\documentclass[12pt,a4paper,final]{iopart}
\usepackage{iopams}
\usepackage{graphicx}
\usepackage[breaklinks=true,colorlinks=true,linkcolor=blue,urlcolor=blue,citecolor=blue]{hyperref}
\usepackage[utf8]{inputenc}

\def \parton#1{\left({{#1}}\right)}
\def \parqua#1{\left[{{#1}}\right]}

\def \derp#1#2{{\partial{#1} \over \partial{#2} }}

\def \mod{\,\mbox{mod}\,}

\def \mod{\,\mbox{mod}\,}
\def \eps{\epsilon}
\def \Aop{{\mathsf A}}
\def \Bop{{\mathsf B}}
\def \Jop{{\mathsf J}}
\def \Pop{{\mathsf P}}
\def \Mop{{\mathsf M}}
\def \xibf{{\mathbf \xi}}
\def \Phibf{{\mathbf \Phi}}
\def \chibf{{\mathbf \chi}}
\def \kbf{{\mathbf k}}
\def \xbf{{\mathbf x}}
\def \ebf{{\mathbf e}}
\def \Toro{{\mathbb T}}
\def \Interi{{\mathbb Z}}
\def \Reali{{\mathbb R}}
\def \Naturali{{\mathbb N}}

\def \Tr{\,\hbox{Tr}\,}

\begin{document}

\title[Errors, correlations, fidelity for noisy Hamiltonian flows.]{Errors, correlations and fidelity for noisy Hamilton flows. Theory and numerical examples.}

\author{G. Turchetti, S. Sinigardi, G. Servizi}
\address{Dipartimento di Fisica e Astronomia, Universit\`a di Bologna and INFN, sezione di Bologna, ITALY.}
\ead{Giorgio.Turchetti@bo.infn.it}
\author{F. Panichi}
\address{Institute of Physics and CASA*, University of Szczecin, Ul. Wielkopolska 15, PL-70-451 Szczecin, POLAND.}
\author{S. Vaienti}
\address{Aix-Marseille Universit\'e, CNRS, CPT, UMR 7332, Marseille, France.}

\begin{abstract}
We analyse the asymptotic growth of the error for Hamiltonian flows due to small
random perturbations. We compare the forward error with the reversibility error, showing their
equivalence for linear flows on a compact phase space.
The forward error, given by the root mean square deviation $\sigma(t)$ of the noisy flow,
grows according to a power law if the system is integrable and according to an
exponential law if it is chaotic.
%The reversibility error is introduced and its
%asymptotic equivalence with the forward error is proved for linear flows.
The autocorrelation and the fidelity,
defined as the correlation of the perturbed flow with respect to the unperturbed one,
exhibit an exponential decay as $\exp\parton{-\sigma^2(t)}$. Some numerical examples
such as the anharmonic oscillator and the H\'enon Heiles model confirm these results.
We finally consider the effect of the observational noise on an integrable system, and show that the decay
of correlations can only be observed after a sequence of measurements and that
the multiplicative noise is more effective if the delay between two measurements is large.

\end{abstract}

\pacs{05.40.Ca, 05.45.Pq, 02.50.Ey}

\noindent{\it Keywords}: Noisy Hamiltonian flow, Fidelity, Reversibility error, Observational noise.

\submitto{\JPA}
\maketitle

\section{Introduction}
The Hamiltonian description of geophysical fluids has been extensively developed \cite{Morrison_4,Morrison_5,Goncharov,Swaters}
and is suited to analyse the conservation laws and to investigate approximation schemes which
preserve the geometric invariants. These methods are also applicable to the Poisson-Vlasov equations for
a collisionless plasma \cite{Morrison_1,Turchetti_5}.
Examples of Hamiltonian models of geophysical fluids are the canonical formulation of Rossby waves 
in a rotating sphere and the system of $N$ vortices.

Finite dimensional Hamiltonian systems
are well understood in the integrable and uniformly hyperbolic
limits, which correspond to an ordered quasiperiodic and
a chaotic evolution \cite{Arnold}. The effect of small
perturbations is amenable to analytical treatment in both situations,
but the case of large perturbations and the transition from
regular to chaotic regimes can only be explored
numerically. In addition the effect of a weak noise and of round off
errors in numerical computations deserves a special attention.
The divergence of orbits and the memory loss rate of a given
system are intimately related and reflect its dynamic behaviour.
The Lyapunov exponents, and related indicators, specify the asymptotic separation rate of
two initially close orbits \cite{Skokos},
whereas the autocorrelation decay measures how fast the evolution loses its memory
of a given initial condition. The asymptotic divergence of nearby orbits, the decay of correlations and the spectrum
of Poincar\'e recurrences are intimately related. In particular
the correlation decay rate is related to the Lyapunov spectrum \cite{Collett}.

When a small  deterministic or stochastic perturbation
is introduced into the system, the perturbed orbit diverges from
the unperturbed one (with the same initial condition) and we call
forward error (FE) their distance at time $t$ or the root mean square deviation
at time $t$, if the perturbation is stochastic. The reversibility error (RE) is the
distance, or the root mean square deviation, from the initial point, of its image
after the evolution forward and backward for the same time interval $t$.
The loss of memory with respect to the reference orbit is given, for instance, by the
classical fidelity \cite{Liverani}, first introduced in quantum systems \cite{Benenti}.

For linear symplectic maps the asymptotic equivalence or RE and FE
was proved in \cite{Panichi} and a relation with the fidelity decay was established.
The fidelity behaviour for maps on the torus and the cylinder
with random perturbations was first analysed in \cite{Marie} and
compared with the perturbation induced by round off errors
\cite{Turchetti_6,Turchetti_7}. The RE analysis was first introduced to investigate the
global effects of round off errors in symplectic maps in \cite{Faranda_1}. 

In this note we examine the asymptotic behaviour of FE and RE for
Hamiltonian flows with a small random perturbation. In the case of linear flows we recover, with
simpler proofs, the results previously obtained for linear symplectic maps.
If the phase space is a compact set such as the torus $\Toro^{2d}$ then
the autocorrelation and fidelity can be computed and their decay rate
is related to the growth of the forward and reversibility errors.

%Starting from the continuity equation we write the evolution of the probability
%density $f_\Toro(x)$ on the torus $\Toro$ (given by the interval $[0,1]$ with identified ends)
%and introduce two different definitions of autocorrelations,
%which coincide when the initial distribution $f(x)$ on $\Reali$ has support on ]0,1, [
%so that $f_\Toro(x)=f(x)$ on$\Toro$ . When a stochastic perturbation
%of small amplitude is introduced,

Denoting with $\sigma(t)$ the FE and RE for a random 
perturbation,  the asymptotic
decay of fidelity is given by $ F(t)\sim e^{-2\pi^2\,\sigma^2(t)}$.
%This result is proved for linear flows. The basic idea is to start from
%the stochastic continuity equation obtaining the Fokker-Planck
%equation by averaging on the noise \cite{Bassi}.
%
% As in the unperturbed case we have two different definitions of
% $C(t)$ and $F(t)$ whose equivalence condition is the same as for the
% unperturbed case. The decay rate depends also on the smoothness of
% the initial condition.
%
For an integrable system defined on the cylinder we prove that the error growth
follows a power law $\sigma(t) \sim \epsilon t^\beta$ where $\beta=3/2$ if
the system is anisochronous, $\beta=1/2$ if it is isochronous. The exponent
$3/2$ is due to the conditional mixing, known as filamentation
in the case of a rotating fluid. This kind of mixing is also responsible for the
algebraic decay of the spectrum of Poincar\'e recurrences \cite{Hu_1}.
These results easily extend to integrable systems of higher dimension.
If the unperturbed system is not skew the result very likely still holds,
as suggested by numerical computations, though the proof is no longer elementary.

The proposed method could be applied to any finite dimensional dynamical system 
with a small additive noise because it is based on the linear equation satisfied
by the Gaussian stochastic process which defines the forward and reversibility errors.
Infinite dimensional extensions might be considered since the stochastic process 
satisfies a linear partial differential equation.

We have analysed numerically the behaviour of the anharmonic oscillator, 
the H\'enon Heiles model, and a system of point vortices \cite{Newton,Aref,Saffman,Luo}.
The results obtained for these models agree with the ones previously
obtained for the standard map and the three body system \cite{Panichi,Turchetti_7}.
Except for the vortex model the Hamiltonian reads $H=T(p)+V(q,t) - \epsilon\, x\,\xi(t)$
where $\xi(t)$ is a white noise and a symplectic integrator was used, 4-th order accurate
for the deterministic component, just as for the three body problem.
For a discussion on numerical stochastic integration see \cite{Mannella}.
In the integrable and chaotic regions the results proved for linear systems are confirmed. 
In the transition regions the results might be interpreted following the model
proposed for Poincar\'e recurrences \cite{Hu_1}.

Random perturbations are introduced by the environment and by observations:
the effect of observational noise in dynamical systems is actively investigated,
see \cite{Faranda_2,McGoff} and references therein.
When the system is integrable the effect of a single observation is small. However a sequence of
observations may cause an exponential decrease of the correlations or the fidelity.
The multiplicative noise changes the signal in a way dependent on the signal itself
and may determine a faster decay of fidelity when the change of phase between two observations is large.

\section{Dynamical systems with additive noise}
In a previous paper we have considered prototype model maps defined on
the torus or the cylinder, whose invariant measure is the normalized
Lebesgue measure.
% For any $A\subset \Toro^1$ the
% invariance of the probability measure $\mu$ is expressed by $\mu(M^{-1}(A))=\mu(A)$.
% For example the Bernoulli map $M(x)=q x\,\mod 1$ where $q\in \Naturali$
% has $q$ inverses and $M^{-1} \equiv \cup_k M^{-1}_k(A)$ where $\mu(M^{-1}_k(A))=q^{-1}\mu(A)$.
% An expanding flow on the Torus $S_t(x)= e^{\beta t} x\mod 1$
% has no unique inverse and the invariance is expressed by $\mu(S_{-t}(A))=\mu(A)$.
%
% The previous analysis on the forward error and fidelity is presented here for the
% corresponding flows perturbed by an additive white noise of small amplitude.
%
We consider first a 1D dynamical system with a stochastic perturbation defined by the Langevin equation
$$
{dx\over dt}= \Phi(x) +\epsilon \xi(t) \eqno(1)
$$
where $x\in \Reali$ and $\xi(t)$ is a white noise.
We denote by $\langle \,\,\,\rangle$ average with respect to all the realizations of the stochastic process
$$
\langle \xi(t)\rangle =0 \qquad \qquad
\langle \xi(t)\xi(t')\rangle =\delta(t-t') \eqno(2)
$$
We denote the deterministic solution with initial point $x_0$ with
$x=S_t(x_0)$ and with $f(x,t)$ the distribution function at time $t$
corresponding to an initial value $f(x_0,0)=f(x_0)$. The
distribution function $f$ for the deterministic system satisfies the
continuity equation
$$
\derp{}{t}f(x,t)+\derp{}{x} (\Phi(x) f(x,t) )=0 \eqno(3)
$$
The fundamental solution of the continuity equation
is given by
$$ G(x,x_0;t)=\delta (x-S_t(x_0)) = {\delta(x_0-S_{-t}(x))\over \eta(S_{-t}(x),t)} \qquad \qquad \eta(x_0,t)= \derp{S_t(x_0)}{x_0} \eqno(4) $$
since the solution $x=S_t(x_0)$ is assumed to have a unique inverse $x_0=S_{-t}(x)$. 
The solution for the distribution function $f(x,t)$ corresponding to an initial distribution $f(x_0,0)=f(x_0)$ reads
$$ f(x,t)= \int _{-\infty}^{+\infty}\, G(x,x_0;t) \,f(x_0) \,dx_0= {f(S_{-t}(x))\over \eta(S_{-t}(x),t) } \eqno(5)
$$
When the noise is present we can still write a stochastic continuity equation
given by (3) where $\Phi$ is replaced by $\Phi(x)+\epsilon\xi(t)$. In this case after
averaging over the noise the distribution function satisfies the Fokker-Planck
equation (see [10]).
$$
\derp{}{t}f(x,t)+\derp{}{x} (\Phi(x) f(x,t) )= {\epsilon^2 \over 2} {\partial^2 \over
\partial x^2}\,f(x,t) \eqno(6)
$$
whose solution is given by
$$
f_\epsilon(x,t)= \int _{-\infty}^{+\infty}\, G_\epsilon(x,x_0;t) \,f(x_0) \,dx_0 \eqno(7)
$$
We denote with $G_\epsilon(x,x_0;t)$ the fundamental solution of the Fokker-Planck
equation, namely its solution with initial condition $G_\epsilon(x,x_0;0) =\delta(x-x_0)$.
In the limit $\epsilon \to 0$ we recover the fundamental solution of equation (3).
We write the solution of the Langevin equation as $x(t)= S_{\epsilon,\,t}(x_0)$ where
$$
S_{\epsilon,\,t}(x_0)= S_t(x_0) +\epsilon \chi(t) = S_t(x_0) +\epsilon \Xi(t) +O(\epsilon^2) \eqno(8)
$$
The stochastic process $\Xi(t)$ is the limit of $\chi(t)$ for
$\epsilon\to 0$ and proves to be Gaussian since it satisfies the linearised equation
$$ {d\,\Xi\over dt} +\Mop(t)\,\Xi= \xi(t) \qquad \qquad \Mop(t) =\Phi'(S_t(x_0)) $$
Indeed $\Xi(t)$ can be written as a convolution of the white noise
$\Xi(t)= \int _0^t \, K(t,s)\,\xi(s)\,ds $.
If $\Phi(x)$ is linear
$$\Phi(x)= \omega +\beta x \eqno(9)$$
then $\chi(t)$ and $\Xi(t)$ are equal. In this case the deterministic component is
$$ S_t( x_0)= e^{\beta t} \,x_0 + \phi(t) \qquad \qquad \phi(t)= \omega\,\,
{e^{\beta t}-1\over \beta} \eqno(10)$$
and the fluctuating component is defined by
$$ \chi(t)= \int_0^t \,\,e^{\beta(t-s)}\,\,\xi(s) ds \qquad \qquad \sigma^2(t)= \,\epsilon^2\, {e^{2\beta t}-1 \over 2\beta} \eqno (11)$$
As a consequence the fundamental solution of the Fokker-Planck equation (6) reads
$$ G_\epsilon(x,t)= { \exp\parton{- {\bigl( \displaystyle x-\langle x \rangle \bigr )^2\over
\displaystyle 2\sigma^2(t)}}
\over \sqrt{2\pi \sigma^2(t) } } \qquad \qquad \langle x(t) \rangle=S_t(x_0)= e^{\beta t}x_0+\phi(t) \eqno(12)$$
We notice that for $\epsilon=0$ the fundamental solution of the Fokker-Planck
reduces to $\delta(x-S_t(x_0))$, namely the fundamental solution of the deterministic
continuity equation (3). We remark also that the root mean square deviation for $\beta>0$ has an exponential
growth $\sigma(t)\sim \epsilon e^{\beta t}$. If $\beta=0$ the motion becomes a translation
and in this case $\sigma(t)= \epsilon \, t^{1/2}$.

\section{Observables on the torus and correlations}
In order to have a chaotic behaviour an exponential divergence of the flow and the phase
space compactness are required. This can be achieved by considering a dynamical system on
the torus $\Toro$. The torus can be seen as the interval $[0,1]$, where the endpoints are identified.
Any periodic function of period 1 defined on $\Reali$ is a dynamic variable on $\Toro^1$.
If $f(x)$ is any function defined on $\Reali$ we may construct a dynamic
variable $f_\Toro(x)$ in the following way
$$ f_\Toro(x)= \sum_{j=-\infty}^{+\infty} \,\, f(x+j) \eqno(13)$$
provided that the decrease of $f(x)$ at infinity is sufficiently fast to insure
the convergence of the series. The function $f_\Toro(x)$ is periodic of period 1 by construction
and therefore can be expanded in a Fourier series
$$ f_\Toro(x)=\sum_{k=-\infty}^{+\infty}\,\,\hat f_k \,e^{2\pi\,i\,k\,x} \eqno(14)$$
whose coefficients are given by
$$ \eqalign{\hat f_k & = \int_0^1 \,e^{-2\pi\,i\,k\,x}\,f_\Toro(x)\,dx = \sum_{j=-\infty}^{+\infty}\,\int_0^1
\,\,e^{-2\pi\,i\,k\,x}\,f(x+j)\,dx \cr
& \cr
& = \sum_{j=-\infty}^{+\infty}\,\int_j^{j+1 }\,\,e^{-2\pi\,i\,k\,(x-j)}\,f(x)\,dx=
\int_{-\infty}^{+\infty}\,\, e^{-2\pi\,i\,k\,x}\,f(x)\,dx \cr} \eqno(15)$$
If $f(x)\in L^1(\Reali)$ its Fourier transform $\hat f(k)$ for $k\in \Reali$ is defined by
$$ \hat f(k)= \int_{-\infty}^{+\infty}\,\, e^{-2\pi\,i\,k\,x}\,f(x)\,dx \eqno(16)$$
and consequently $\hat f_k=\hat f(k)$ for $k\in \Interi$.

If $f(x,t)$ is the solution of the continuity equation on $\Reali$ the corresponding solution
on $\Toro$ is given by
$$ \eqalign{ f_\Toro(x,t)= \sum_{j=-\infty}^{+\infty} \,\, f(x+j,t)= & \sum_{j =-\infty}^{+\infty}\,\,\int_{-\infty} ^{+\infty}
\delta(x+j-S_t(x_0)) \,f(x_0)\,dx_0 =\cr
& \cr
&= \,\int_{-\infty} ^{+\infty}\,G_\Toro(x,x_0,t)\,f(x_0) \,dx_0 \cr} \eqno(17) $$
By $G_\Toro(x,x_0,t)$ we denote the fundamental solution on $\Toro^1$
whose Fourier expansion is given by
$$ G_\Toro (x,x_0;t)= \sum_{j =-\infty}^{+\infty}\,\delta(x+j-S_t(x_0)) \, =
\sum_{k=-\infty}^{+\infty}\,\, e^{2\pi\,i\,kx} \, e^{-2\pi\,i\,k\,\,S_t(x_0)} \eqno(18)$$
By replacing (18) into (17) we obtain the Fourier expansion of $f_\Toro(x,t)$ according to
$$ f_\Toro(x,t)=\sum_{k=-\infty}^{+\infty} \,\, \hat f_k(t) \,e^{2\pi\,i\,k\,x}\,
\qquad\qquad \hat f_k(t) = \int_{-\infty}^{+\infty}\,dx_0 \,e^{-2\pi\,i\,k\, S_t(x_0)}\, f(x_0)
\eqno(19) $$
We introduce also the following coefficients
$$ f_k(t)= \int_0^1\,dx_0 \,e^{-2\pi\,i\,k\, S_t(x_0)}\, f_\Toro (x_0) \eqno(20) $$
Notice that for $t=0$ we have $\hat f_k(0)=\hat f_k$.
The new time dependent coefficients differ from the Fourier components unless the evolution is just
a translation $S_t(x)= x_0+\omega t$ or $f(x)$ has support on $[0,1]$ vanishing elsewhere. In this case
% $f_\Toro(x)= f(x)$ in $[0,1]$ and
%
$$ f_k(t)=\hat f_k(t)    $$
and the following equality holds
$$ \int_0^1 f_\Toro(x) \,f_\Toro(x,t) \,dx = \int_0^1 \, f_\Toro(x_0) \,f_\Toro(S_t(x_0))\,dx_0 \eqno(21)$$

\subsection{Correlations}
We introduce two distinct definitions of the correlation on the torus according to
$$ \hat C(t)= \int_0^1 \,f_\Toro(x) \,f_\Toro(x,t))\,dx -f_0^2 = \sum_{k\not =0} \,\hat f_{-k} \hat f_k(t) \eqno(22)$$
see equation (19)
$$ C(t)= \int_0^1 \,f_\Toro(x_0) \,f_\Toro(S_t(x_0))\,dx_0 -f_0^2 = \sum_{k\not =0} \,\hat f_{-k} f_k(t) \eqno(23) $$
see equation (20).
These definitions are equivalent when $S_t$ is a translation or $f(x)$ has support on $[0,1]$.
In general we can write $C(t)$ according to
$$ C(t)= \sum_{(k',k) \not=(0,0) } \hat f_{k'}\,A_{k', -k}\,\hat f_{ -k }
\qquad \qquad A_{k',-k}= \int_0^1 \,dx_0\,\,e^{2\pi\,i(k'\,x_0 - k\,S_t(x_0)\,)} \eqno(24) $$
When the vector field $\Phi(x)$ is linear, see equation (9), and the flow is linear, see equation (10), then $C(t)=\hat C(t)$
for a sequence of times $t=t_m$ where $t_m=\beta^{-1}\, \log m$ for $m\in \Interi$. Indeed from equations (19), (10) and (16) we have
$$ \hat f_k(t)= e^{-2\pi\,i\,k\, \phi(t)} \,\hat f(ke^{\beta t}) \eqno(25) $$
The coefficients $f_k(t)$, according to equations (23) and (24), are given by
$$ f_k(t)= \sum _{k'=-\infty}^{+\infty}\, \hat f_{k'} \, A_{k',-k}\, \eqno(26) $$
where $A_{k',-k}$ is given by
$$A_{k',-k}= %e^{-2\pi\,i\,k\phi(t) }\,\int_0^1 dx_0\,e^{2\pi\,i\,x_0(k'-e^{\beta t}k}=
e^{-2\pi\,i\,k\phi(t) }\,
\cases{ e^{\pi i (k'-k\,e^{ \beta t} ) } \sin(\pi(k'-ke^{\beta t})/(\pi(k'-k\,e^{\beta t})) & if $e^{\beta t}\not = m$ \cr & \cr
\delta_{k', mk} & if $e^{\beta t} = m$ \cr } \eqno(27)$$
As a consequence, for $t=t_m$, from (26) and (27) we have
$$ f_k(t_m)=\hat f_k(t_m)=e^{-2\pi\,i\,k\phi(t_m) } \hat f_{km} = e^{-2\pi\,i\,k\phi(t_m) } \hat f_{ke^{\beta t_m}} \eqno(28) $$

\section{Noisy systems on the torus: correlations and fidelity}
For a noisy system the evolution of a given initial distribution is given by equation (7) and if the system is defined on
the torus $\Toro$ the evolution of the corresponding distribution $f_\Toro(x)$ defined by (13) is a periodic function in $x$ which can be
expanded in a Fourier series according to
$$ f_{\epsilon,\,\Toro}(x,t) = \int _{-\infty}^{+\infty}\, G_{\epsilon\,\Toro}(x,x_0;t) \,f_0(x_0) \,dx_0
= \sum_{k=-\infty}^{+\infty}\,\hat f_{\epsilon,\,k}(t) \,e^{2\pi\,i\,kx} \eqno(29)$$
where $G_{\epsilon\,\Toro}(x,x_0;t)$ denotes the fundamental solution of the Fokker-Planck equation on the torus
and $f_\Toro(x,t)$ the distribution function on the torus at time $t$
$$ G_{\epsilon\,\Toro}(x,x_0;t)=\sum_{j=-\infty}^{+\infty}\,\,G_{\epsilon}(x+j,x_0;t)
\qquad\qquad f_{\eps,\,\Toro}(x,t)=\sum_{j=-\infty}^{+\infty} f_\eps(x+j,t) \eqno(30) $$
Even in this case we introduce two distinct definitions of correlations as an obvious extension of the noiseless case
$$ \hat C_\epsilon(t)= \int_0^1 \,f_\Toro(x) \,f_{\epsilon,\,\Toro}(x,t)\,dx -f_0^2 =
\sum_{k\not=0} \, \hat f_{-k}\,\hat f_{\epsilon,\,k}(t) \eqno(31) $$
and
$$ C_\epsilon(t)= \int_0^1 \,f_\Toro(x_0) \,\langle f_{\Toro}\bigl(\,S_{\epsilon,\, t}(x_0)\,\bigr)\rangle \,dx_0
-f_0^2 = \sum_{k\not=0}\, \hat f_{-k}\, f_{\epsilon,\,k}(t) \eqno(32) $$
where
$$ f_{\epsilon,k}(t)= \int_0^1 \, \langle e^{-2\pi\,i\,k \,S_{\epsilon,t}(x_0) }\rangle
\,f_\Toro(x_0)\,dx_0$$
The fidelity gives the correlation between a dynamic variable $f(x)$ evaluated along the perturbed orbit at time $t$ and $f(x)$
evaluated along the unperturbed orbit at the same time. We have a unique definition of fidelity given by
$$\eqalign{F_\epsilon(t) &= \int_0^1 \,f_\Toro(S_t(x_0)) \,\langle f_{\Toro}\bigl(\,S_{\epsilon,\, t}(x_0)\,\bigr)\rangle \,dx_0 -f_0^2 = \cr
& = \sum_{(k,k')\not =(0,0)} \,
\hat f_{-k} \hat f_{k'} \,\,\int_0^1 \,dx_0 \langle e^{2\pi\,i\, (k'\,S_{\epsilon,\, t}(x_0)-kS_t(x_0)} \rangle} \eqno(33) $$

\subsection{Linear systems}
For a linear Langevin equation, namely for $\Phi(x)=\beta x +\omega$, explicit expressions can be obtained. Indeed,
to evaluate $\hat f_{\eps,\,k}(t)$, we expand $G_{\eps\,\Toro}(x,x_0;t)$ in Fourier series, taking (8) and (11) into account. The result is
$$\eqalign{ G_{\eps, \,k}(x_0,\,t)= \int_0^1\, dx \,e^{-2\pi\,i\,kx}\,\,G_{\epsilon\,\Toro}(x,x_0;t) & =\int _{-\infty}^{+\infty}\,
e^{-2\pi\,i\,kx}\,{ \exp\parton{- {\bigl( \displaystyle x- S_t(x_0) \bigr )^2\over
\displaystyle 2\sigma^2(t)}} \over \sqrt{2\pi \sigma^2(t) } } \cr
& \cr
& = e^{-2\pi ik\,S_t(x_0)}\,e^{-2\pi^2\,\sigma^2(t)} \cr} \eqno(34) $$
The Fourier components of $f_{\eps\,\Toro}(x,t)$ are given by
$$ \hat f_{\epsilon,\,k}(t) =e^{-2\pi^2\,\sigma^2(t)}\,\,\int _{-\infty}^{+\infty}\,\,e^{-2\pi ik\,S_t(x_0)}\,f(x_0) \,dx_0=
e^{-2\pi^2\,\sigma^2(t)}\,\,\hat f_k(t) \eqno(35) $$
In order to evaluate $f_{\eps,\,k}(t)$ we recall that for a linear Langevin equation (1) where $\Phi(x)$ is given by (9)
the solution $S_{\eps,\,t}(x_0)$ is given by (8) and (11) so that
$$ \langle e^{-2\pi\,i\,kS_{\epsilon,\, t}(x_0) }\rangle= e^{-2\pi\,i\,k\,S_t(x_0)}\langle e^{-2\pi\,i\,k\,\epsilon{\chi(t)}}\rangle=
e^{-2\pi\,i\,k\,S_t(x_0)}\,e^{-2\pi^2k^2 \sigma^2} \eqno(36) $$
As a consequence taking equation (16) into account we can write
$$ f_{\epsilon,\,k}(t)= e^{-2\pi^2k^2 \sigma^2} f_{k}(t) \eqno(37) $$
The explicit expression for the correlation is
$$ \hat C_\epsilon(t) =\sum_{k\not=0} \,\hat f_{-k}\,\hat f_{\,k}(t) \, e^{-2\pi^2 k^2 \sigma^2(t)} \eqno(38) $$
and
$$ C_\epsilon(t)= \sum_{(k,k')\not =(0,0)}\, \hat f_{k'}\,A_{k',-k} \, \hat f_{-k} \, ,e^{-2\pi^2\,k^2\, \sigma^2(t) } \eqno(39)$$
In this case the fidelity has the following expression
$$ F_\epsilon(t)= \sum_{(k,k')\not =(0,0)} \, \hat f_{-k}\, \hat f_{k'}\, \int_0^1\,dx_0\, e^{2\pi\,i\,(k'-k)\,S_t x_0}\,\,\,\,\, e^{-2\pi^2 \sigma^2(t) } \eqno(40) $$

\section{Model systems}
We consider first three different prototype models

\noindent
i) { \bf Translations on the torus $\Toro^1$.}

The field is $\Phi(x)=\omega$ and $S_t(x_0)=x_0+\omega t $ and the Fourier coefficients are
$$ f_k(t)=\hat f_k(t)= e^{-2\pi\,i\,k\,\omega t} \,\hat f_k \eqno(41)$$
The correlations do not decay.
\par \noindent
When the noise is introduced the root mean square deviation with respect to the unperturbed trajectory is $\sigma= \epsilon \,t^{1/2}$
and the Fourier coefficients are given by
$$ f_{\epsilon\,k} (t)=\hat f_{\epsilon\,k}(t)= e^{-2\pi\,i\,k \,\omega t} \,\hat f_k\, e^{-2\pi^2 \epsilon^2 t} \eqno(42) $$
The decay of correlations and fidelity is exponential with decay time $\tau=(2\pi^2\epsilon^2)^{-1} $.
The difference is that the correlations oscillate and decay, while the fidelity goes to zero without oscillations, due to the
absence of the phase factor.

\noindent
ii) {\bf Anisochronous translations on the cylinder}

This is a skew system. Letting $\xbf=(x,y)\in {\mathbb C}= \Toro\times {\mathbb I}$ be the phase space coordinates, $\Phibf=(y,0)$ the
vector field and $S_t(\xbf_0)= ( x_0+y_0 t, y_0)$ the flow, for an observable $f(x)$ to which corresponds $f_\Toro(x)$ on the torus
we have
$$ \hat f_k(t)= f_k(t)= e^{-i\,k\,y_0 t} \, \hat f_k \eqno(43)$$
Even though $f_\Toro$ does not depend on the torus label $y_0$ the coefficient $f_k(t)$ depend on it according to the previous equation.
The correlation is computed by averaging on the interval ${\mathbb I}=[0,a]$ to which $y_0$ belongs.
Taking account of the normalization factor $a^{-1}$ the correlation reads
$$C(t)=\hat C(t)= \sum_{k\not=0} \,|f_k|^2 \,{1\over a} \,\int_0^a\,e^{-2\pi\,i\, k\,y_0\, t } \,dy_0=
\sum_{k\not=0} \,|f_k|^2 \,\ e^{-\pi\,i\,k\,a\, t} \,\,
{\sin(\pi \,k\, a\,t)\over \pi\, k \, a\,t } \eqno(44)$$
as a consequence supposing that $|f_k|<A/|k| $ we have
$$|C(t)|<{3A^2\over \pi a} \,{1\over t} \eqno(45)$$
and the decay follows a power law. This slow decay is due to the conditional mixing, known in physics as the filamentation phenomenon,
which occurs because the frequency on the tori has a continuous and monotonic variation.
For an observable $f(x,y)$ such that the average on $x$ is constant,
the result still holds provided that the Fourier coefficients $\hat f_k(y)$ are $C^1$ and decay faster than $1/y$ at infinity.

Any autonomous 1D system in the region around a stable fixed point delimited by the separatrices is defined,
using action angle variables $(\phi,\jmath)$, by the vector field $\Phibf=( \omega(j), 0)$. If $\omega(j)$ is monotonic,
introducing the new variables $x=\phi/(2\pi)$ and $y=\omega(\jmath)$ we are back to the case considered above.

When a stochastic perturbation is introduced the vector field becomes $\Phibf(x,y)= \bigl (\,y+ \epsilon \xi_x(t), \epsilon \xi_y(t) \, \bigr)$ where
$\xi_x(y)$ and $\xi_y(t) $ are two independent white noises. In this case the stochastic flow $S_{\epsilon,\,t}(x_0,y_0)$
is given by
$$ x=x_0+y_0 t\,+\epsilon w_y^{(1)}(t)+ \epsilon w_x(t) \qquad \qquad y=y_0+ \epsilon w_y(t) \eqno(46)$$
where $w(t)=\int_0^t \,\xi(s)\,ds$ denotes a Wiener noise and $w^{(1)}(t)=\int_0^t \,w(s)\,ds$.
Recalling that $\langle w_x^2(t)\rangle=t$, $\langle( w_y^{(1)}(t))^2\rangle=t^3/3$,
$\langle w_x(t)w_y^{(1)}(t) \rangle=0$, the root mean square deviations for $x$ and $y$ are given by
$$ \sigma_x=\epsilon \parton{ t+ {t ^3\over 3} }^{1/2} \qquad \qquad \sigma_y=\epsilon \, t^{1/2} \eqno(47)$$
The correlation on the cylinder for the observable $f_\Toro(x)$ is given by
$$ C(t)= \sum_{k\not =0 } |f_k|^2\, \parqua{ e^{- \pi\, i\,k\, a\, t} \,\,
{\sin(\pi \,k\, a\,t)\over \pi\, k \, a\,t } } \, e^{-2\pi^2\,k\, \eps^2(t + t^3/3)} \eqno(48)$$
The fidelity has the same expression where the term within the square parenthesis is
replaced by 1. If the frequency is the same on each torus, or if the $y$ variable is not affected by noise, the
decay of correlations or fidelity is much slower since the cubic term in the exponential is absent.

\noindent
iii) {\bf Expanding flow}

Given the vector field $\Phi(x)=\beta x$, the flow is $S_t(x_0)= e^{\beta t} x_0 $
and the Fourier coefficients are given by
$$ \hat f_k(t)= \int_{-\infty}^{+\infty} \, f(x)\,e^{2\pi\,i\,k\,x_0e^{\beta t}}= \hat f(k\,e^{\beta t}) \eqno(49)$$
If $f(x)\in L^2$ then $\hat f(k) \in L^2$ and consequently $|\hat f(k)|$ decreases to $0$ for $|k| \to \infty$.
If $|\hat f_k| \le A\,|k|^{-1}$ then $\hat f(ke^{\beta t})$ decreases exponentially fast to zero with $t$ and the following estimate holds
$$ |\hat C(t) | \le 2A^2\,e^{-\beta t} \,\sum_{k=1}^\infty \, {1\over k^3} \le 3A^2 \,e^{-\beta t} \eqno(50) $$
The Fourier coefficients entering the other definition of correlation are, for $e^{\beta t}\not \in \Naturali$,
$$ f_k(t)= \int_0^1 \,f_\Toro(x_0) \, e^{ - 2\pi\,i\,k \,x_0e^{2\beta t} } \,dx_0= \sum _{k'=-\infty}^{+\infty} \, f_{k'}
e^{ + \pi\,i\, (k'-ke^{\beta t}) }\,{ \sin \pi(k'-ke^{\beta t}) \over k'-ke^{\beta t} } \eqno(51)$$
If $e^{\beta t}=m \in \Naturali$ then letting $t_m=\beta^{-1} \, \log m $ we have
$ \hat f_k(t_m) = f_k(t_m)= \hat f_{m k} $ where $ |\hat f_{mk}|\le \,e^{-\beta t_m}\,A/|k|$.
If $f(x)$ has a compact support on $[0,1]$ the two definitions agree for any $t\in \Reali$.

When the noise is present the root mean square deviation is given by
$$ \sigma= \parton{e^{2\beta t}-1 \over 2 \beta}^{1/2} \eqno(52)$$
The correlations consequently are given by (38) and (39) , the fidelities by (40) where $\hat f_k(t)$ and $ f_k(t)$ are given by equations (49) and (51) respectively.

The explicit expressions of correlations for $e^{\beta t} \not \in \Naturali$ is

$$ \hat C_\epsilon(t)= \sum_{k\not=0} \,\hat f_{-k}\, \hat f(ke^{\beta t})\,e^{-2\pi^2 k^2\sigma^2(t)} \eqno(53)$$
and
$$ C_\epsilon(t)= \sum_{(k,k')\not=(0,0)} \, f_{-k}\,f_{k'}\,\, e^{\pi\,i\,(k'-ke^{\beta t})}\,\,{\sin\pi((k'-ke^{\beta t}) \over \pi(k'-ke^{\beta t})} \,
e^{-2\pi^2 k^2\sigma^2(t)} \eqno(54)$$
The explicit expression for the fidelity reads
$$ F_\epsilon(t)= \sum_{(k,k')\not=(0,0)} \, f_{-k}\,f_{k'}\,\, e^{\pi\,i\,(k'-k)e^{\beta t}}\,\,{\sin\pi((k'-k)e^{\beta t} \over \pi\,i(k'-k)e^{\beta t}}\,\,e^{-2\pi^2 k^2\, \sigma^2(t)} \eqno(55)$$
For $e^{\beta t}=m \in \Naturali$ we have
$$ \hat C_\epsilon(t_m)= C_\epsilon(t_m)=\sum_{k\not=0} \,\hat f_{-k}\, \hat f_{mk}\, e^{-2\pi^2 k^2\sigma^2(t_m)}
\qquad F_\epsilon(t_m)= \sum_{k\not =0}\,|f_k|^2\,\,e^{-2\pi^2 k^2\sigma^2(t_m)} \eqno(56)$$
%
%
%$$ F(t)= \sum_{(k,k')\not=(0,0)} \, f_{-k}\,f_{k'}\,\, e^{\pi\,i\,(k'-k)e^{\beta t}}\,\,{\sin\pi((k'-k)e^{\beta t} \over \pi\,i(k'-k)e^{\beta t}}\,\,\,\,e^{-2\pi^2 k^2\sigma^2(t)} \eqno(55)$$

\section{Reversibility error and related fidelity}
The reversibility error is the distance from the initial point of its forward evolution up to time $t$ and backward for the
same time interval. Any autonomous flow is reversible since the flow $S_t$ has the group property $S_{-t}\circ S_t=I$.
This property is lost when the system is stochastically perturbed. In this case the local errors
$\epsilon \xi(\pm t)$ at times $\pm t$ are independent and we have $S_{\epsilon,-t}\circ S_{\epsilon,t}\not =I$.

We define a fidelity associated to the reversibility error according to
$$ F_{\epsilon\,R}(t)= \int_0^1 \,dx_0 \,f_\Toro(x_0)\, \langle f_\Toro(S_{\epsilon,-t}\circ S_{\epsilon,t}(x_0)) \rangle - f_0^2 \eqno(57) $$
The FE and the RE are related, and for linear flows we prove they are equal up to a factor $\sqrt{2}$.

We consider first the translation perturbed with a white noise whose flow is $S_{\eps,\,t }\,x_0= x_0+\omega t +\epsilon w(t)$ so that
$S_{\eps,\,-t }\circ S_{\eps,\,t }\,x_0= x_0+\eps(w(t)+w(-t))$. The reversibility error is given by the root mean square deviation of this process
$\sigma_R(t)= \epsilon \sqrt{2} \,t^{1/2} $, since $\langle w(t)w(-t)\rangle=0$.
If the flow is expanding, namely $\Phi_{\eps}= \beta x +\eps \xi(t)$ then we have
$$S_{\eps,\,-t }\circ S_{\eps,\,t }\,x_0= x_0+ \eps \chi_R(t) \qquad \qquad \chi_R(t)= e^{-\beta t}\chi(t)+\chi(-t) \eqno(58) $$
As a consequence from
$$ \chi_R(t)= \int_0^t e^{-\beta s }\,(\xi(s)-\xi(s-t) ) ds $$
we obtain the reversibility error
$$ \sigma_R(t)= \eps \langle \chi_R^2(t) \rangle^{1/2 }=
\eps \sqrt{2}\, {1-e^{-2\beta t}\over 2 \beta } \eqno(59)$$
The reversibility error $\sigma_R(t)$ is equal to the forward error $\sigma(t)$ (times $\sqrt{2}$)
but for the system whose vector field is $\Phi(x)=-\beta x$ rather than $\Phi(x)=\beta x$.
The forward flow is expanding, the reverse flow is contracting, but RE for the expanding flow is equal to the FE
for the contracting flow. To recover the equivalence it is sufficient to consider the flow on the torus $\Toro^2$
defined by the following noisy vector field $\Phi_\epsilon= (-\beta x +\epsilon\xi_x(t), \beta y+ \epsilon \xi_y(t))$
where $\xi_t(t)$ and $\xi_y(t)$ are independent white noises. A simple computation shows that in this case
$\sigma_R(t)=\sqrt{2}\,\sigma(t)$. The deterministic flow in this case is area preserving.
More generally linear Hamiltonian flows exhibit the same equivalence asymptotically.

\subsection{Quadratic Hamiltonians}

We consider the general case of a quadratic Hamiltonian $H(\xbf)={1\over 2}\,\xbf\cdot \Bop\xbf$ where $\xbf\in \Reali^{2d}$ for an arbitrary number
of degrees of freedom $d$ and $\Bop$ is a symmetric matrix. The equations of motion and the flow are
$$ {d\xbf\over dt}= \Aop\equiv \Jop \Bop \qquad \qquad \Jop = \pmatrix{0 & I \cr -I & 0 \cr} \qquad\qquad S_t 	\xbf_0= e^{\Aop t}\xbf_0 \eqno60)$$
where the matrix $e^{\Aop t} $ is symplectic namely $e^{t\Aop}\Jop \,e^{t\Aop^T} = \Jop $. When the system is stochastically perturbed the equations of motion and the stochastic flow become
$$ {d\xbf \over dt}= \Aop \xbf + \eps\,\xibf(t) \qquad \qquad S_{\eps,\,t}\xbf_0= e^{\Aop t }\xbf_0+ \epsilon \chibf(t)
\qquad \qquad \chibf(t)= \int_0^t e^{\Aop(t-s) }\,\xibf(s)\,ds \eqno(61)$$
The RE is computed starting from the reversed flow
$$ S_{\eps,\,-t}\, S_{\eps,\,t}\,\xbf_0= \xbf_0+ \chibf_R(t) \qquad \quad \chibf_R(t) = \eps\bigl [e^{-\Aop t} \chibf(t) +\chibf(-t) \bigr ]
= \int_0^t e^{- s \Aop}( \,\xibf(s) - \xibf(s-t) ) \,\,ds \eqno(62)$$
and explicitly reads
$$ \sigma^2_R(t)= \eps^2 \, \left \langle \, \left \Vert \chibf_R(t) \right \Vert ^2\,\, \right \rangle = 2\eps^2 \,\int_0^t\,ds \, \Tr\parton{e^{-s\,\Aop} \, e^{-s\,\Aop^T} } \eqno(63) $$
The forward error $\sigma(t)$ is defined by
$$ \sigma^2(t)= \eps^2 \, \left \langle \, \left \Vert \chi (t) \right \Vert ^2\,\, \right \rangle = \eps^2 \,\int_0^t\,ds \, \Tr\parton{e^{s\,\Aop} \, e^{s\,\Aop^T} } \eqno(64) $$
It is not hard to prove that $\sigma(t)$ and $\sigma_R(t)/\sqrt{2}$ are asymptotically equal taking into account
the property of symplectic matrices. Indeed $e^{\Aop t}$ and $e^{-\Aop t}$ have the same eigenvalues.
\par\noindent
The fidelity $F_R(t)$ for a noisy reversed flow on the torus $\Toro^{2d}$ is defined by
$$ F_R(t)= \int _0^1 \,d\xbf_0 \,\,f_{\Toro^{2d}}\, (\xbf_0) \,\,f_{\Toro^{2d}}\, ( S_{\eps,\,-t} S_{\eps,\,t}\,\xbf_0) \,
-\parton{\int _0^1 \,d\xbf_0 \,\,f_{\Toro^{2d}}(\xbf_0}^2 \eqno(65)$$
The reversed flow is given by equation (62) where the global error $\epsilon\chi_R(t)$ is defined. Expanding $f_{\Toro^{2d}}(\xbf)$ in
a Fourier series we find
$$ \eqalign{ F_R(t)= \sum_{(\kbf,\kbf')\not = ({\bf 0},{\bf 0})} \,f_{-\kbf}\,f_{\kbf'} \,\, \int_0^1\,dx_{0\,1}\,\cdots\, & \int_0^1\,dx_{0\,2d}
\,\, e^{-2\pi \,i\,( \kbf'-\kbf) \cdot\xbf_0}\,\langle e^{2\pi \,i\, \kbf' \,\eps\,\chibf_R(t) } \rangle= \cr
&= \sum_{(\kbf,\kbf')\not = ({\bf 0},{\bf 0})} \,f_{-\kbf}\,f_{\kbf'} \,e^{-4\pi^2\,\kbf \cdot \Sigma^2(t)\, \kbf} \cr } \eqno(66)$$
where
$$ \Sigma^2(t)= \eps^2 \int _0^t ds \,\,e^{ - \Aop s}\,e^{-\Aop^Ts} \eqno(67)$$
We recall that the square of the reversibility error is expressed by
$$ \sigma_R^2(t)= \eps^2\, \Tr\bigl (\Sigma^2(t)\bigr ) \eqno(68)$$
If all the eigenvalues are complex of unit modulus then $e^{s \Aop }$ is conjugated to an orthogonal matrix.
If $e^{s \Aop }$ is orthogonal $\sigma^2(t)= 2d \,\eps^2 \,t$ supposing the phase space is $\Reali^{2d}$.
If $A$ has real eigenvalues $\pm \lambda_i$ ordered in an increasing sequence $-\lambda, -\lambda_2,\ldots, \lambda_2, \lambda$
then $e^{tA}$ is conjugated to the diagonal matrix $ e^{t\Lambda}=\hbox{diag}\,(e^{\lambda t}, \,e^{\lambda_2\,t}, \ldots, e^{-\lambda_2 t}, e^{-\lambda t} )$. Supposing for simplicity that $\Aop=\Lambda$ is diagonal (in the general case only trivial constant factors appear) we obtain that asymptotically
$$ \sigma(t)\simeq \epsilon e^{\lambda t} \qquad \qquad \sigma_R(t)\simeq \,\sqrt{2} \,\eps\, e^{\lambda t} \qquad \qquad \kbf\cdot \Sigma^2_R(t) \kbf=k_1^2 \,\eps^2 \, e^{2\lambda t} \eqno(69) $$
Also in this more general case the asymptotic decay rate of fidelity is related to the asymptotic growth of the error.

\section{Models}
We give some examples of application of the previous results. To this end we consider as integrable system the anharmonic oscillator and as non integrable
system the H\'enon Heiles Hamiltonian. The Hamilton's equations are integrated by using symplectic 4-th order integrators, based on the splitting method.
Finally we consider the system of vortices for which explicit symplectic integrators are not available: we use then a Runge-Kutta integrator.
The restricted 3 body problem has been discussed elsewhere \cite{Panichi} just as native discrete systems expressed by symplectic maps \cite{Turchetti_6}.

Denoting the time step by $\Delta t= T/n_s$, where $T$ is a characteristic time of the system, and using fourth order methods, we insure that the local integration error is close to the round off choosing $n_s=10^4$. In addition symplectic integrators insure that the error on the invariants
has a null average growth. We consider also the Lyapunov error defined as the distance at time $t$ from the reference orbit of an orbit whose initial
point has an error of amplitude $\epsilon$. 
We have checked that for integrable systems the forward error due to a stochastic perturbation
of amplitude $\epsilon$ grows as $\sigma(t)\sim \epsilon t^{3/2}$ just as the reversibility error up to a factor $\sqrt{2}$. However when
the system is isochronous the growth is $\sigma(t) \sim \epsilon t^{1/2}$. We have checked that decay of correlations and fidelity follows
an exponential law according to $C(t)\sim F(t) \sim e^{-2\pi^2 \sigma^2(t)}$.
For non integrable systems in the regions of regular motion,
the error growth follows a power law with exponent $3/2$, whereas in the chaotic regions
the growth is exponential $\sigma (t) \sim \epsilon e^{\lambda t}$, where $\lambda$ is the maximum Lyapunov exponent.

\subsection{The anharmonic oscillator}
The Hamiltonian is given by
$$ H_\eps= {\omega \over 2}(x^2+p^2)+{\eta\over 4} x^4 - \epsilon p \xi(t)= \omega \jmath+ \eta\,\, \jmath^2 \cos^4 \theta +\eps (2\jmath)^{1/2} \sin \theta \,\,\xi(t) \eqno(70)$$
The transformation to new action angle variables $(\Theta,J) $ allows to render $ H$ integrable up to a remainder of
order $\lambda^2$ and neglecting this remainder as well as terms of order $\lambda \eps$ the Hamiltonian reads
$$ H=\omega J + \eta \, {3\over 8}{J^2\over \omega^2} \,+ \, \epsilon (2J)^{1/2} \, \sin \Theta \,\,\xi(t) \eqno(71)$$

\begin{figure}[htb!]
\centering
\includegraphics[width=8cm,height=7 cm]{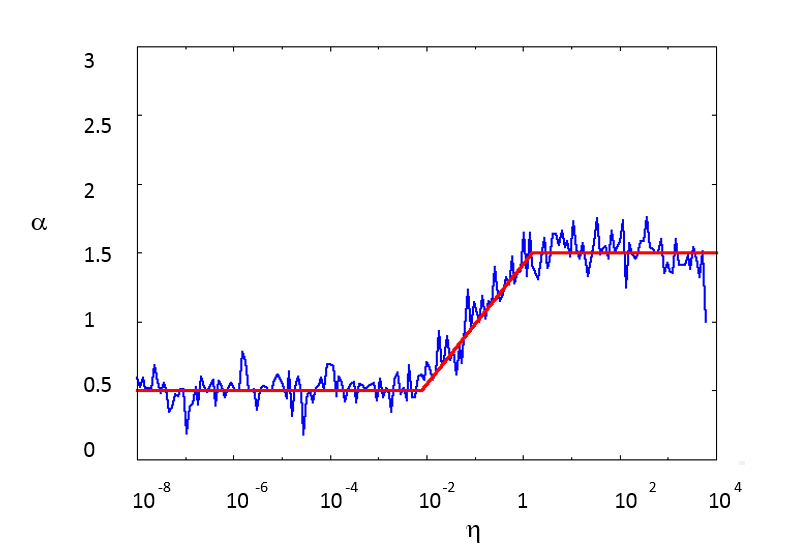}
\caption{\label{Figure_1} Plot of the exponent $\alpha$ for the fit $\sigma(t)= C\,\epsilon\,n^\alpha$ to the forward error of
the anharmonic oscillator with a stochastic perturbation of amplitude $\epsilon$ as a function of the nonlinearity strength $\eta$. The noise is additive in Cartesian coordinates, according to equation (70).
The red line is drawn just to guide the eye. When $\eta \ll1$ the system is almost isochronous and $\alpha=1/2$,
when $\eta \gg1 $ the system is strongly anisochronous and $\alpha=3/2$, but in the intermediate region the exponent varies continuously between these two values.}
\end{figure}

In these new coordinates the noise become multiplicative. A much simpler model to solve analytically is the one in which the
noise is additive
$$ H=H_0(J ) \,- \Theta\,\epsilon \,e_J\, \xi_J(t)+ J\, \epsilon \,e_\Theta\,\xi_\Theta(t) \qquad \qquad H_0(J)=\omega J +\eta\, {3\over 8} \,{J^2\over \omega^2} \eqno(72)$$
where $\ebf=(e_\Theta,e_J)$ is a unit vector and $\xi_\Theta$, $\xi_J(t)$ are independent white noises.

In this case the equations of motion can be solved explicitly by retaining only the first order in $\eps$ and
for initial conditions $J(0)=J_0,\, \Theta(0)=\Theta_0$ read
$$ \Theta= \Theta_0+\Omega(J_0)t+ \eps \Omega'(J_0) \,e_J\, w_J^{(1)}(t) \,+\eps\,e_\theta\,w_\theta(t)
\qquad \qquad J(t)=J_0+\eps \,e_J\, w_J(t) \eqno(73) $$
where $\Omega =dH_0/dJ$ and $\Omega'=d\Omega/dJ$ . As long as the distance on the cylinder agrees with the Euclidean distance we can write for the forward error
$$ \sigma(t)= \eps\parqua{t +e_J^2 \,{ \Omega' }^2 (J_0) \, t^3/3} ^{1/2} \eqno(74) $$
The figure shows the result of a fit to the exponent $\alpha$ for power law growth of the forward error $\sigma(t)= C \epsilon \, t^\alpha$
as a function of the anisochronicity strength $\eta$. The function $\alpha(\eta)$ varies between $1/2$ and $3/2$ with a rather sharp transition
in agreement with theoretical estimate (74).
\begin{figure}[htb!]
\centering
\centerline{
\includegraphics[width=8cm,height=7 cm]{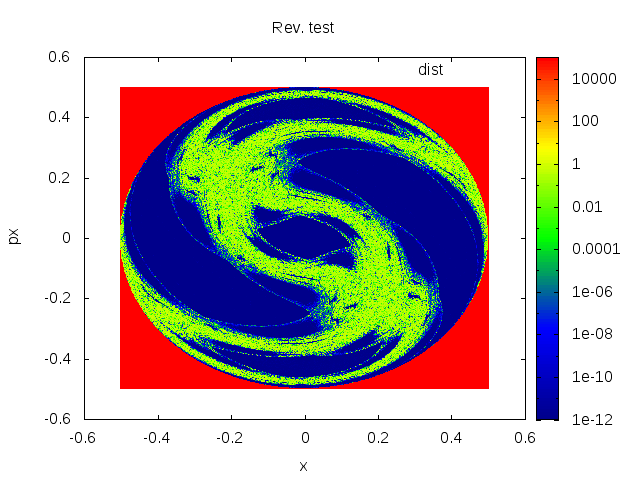}
\includegraphics[width=8cm,height=7 cm]{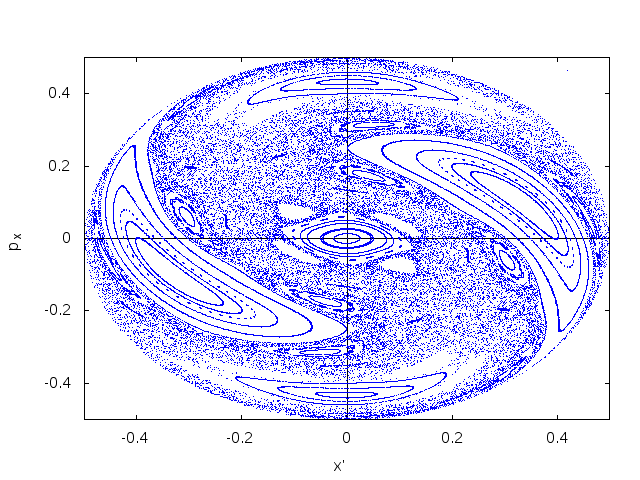}
}
\caption{\label{figureone}H\'enon-Heiles non integrable Hamiltonian $H_-$ given by equation (75). 
Left side: colour plot of the reversibility error due to round off for $H_-=1/8$ and initial conditions 
on a grid of points in the $(x,p_x)$, with $y=0,p_y>0$ and $t=100$ dynamic periods $T=2\pi$.
Right figure: phase space plot for the projection into the $(x, p_x)$ phase plane of the orbits in the Poincar\'e section
$y=0, p_y>0$ }
\end{figure}
\subsection{The H\'enon Heiles Hamiltonian}
This model describes the motion of an elliptical galaxy and the Hamiltonian reads
$$ H_\pm={1\over 2}(p_x^2+p_y^2) +{1\over 2}(x^2+y^2) +x^2y\pm{1\over 3} y^3 \eqno(75)$$
The Hamiltonian $H_-$ is not integrable, whereas the model $H_+$ is integrable. For the first one we have considered the
reversibility error due to the round off. In Figure 2 we compare the colour plot for the reversibility error with the phase
portrait. Both refer to the Poincar\'e section $y=0,\dot y>0$

\subsection{The N vortex model}
For inviscid plane fluid the dynamics of point vortices provides a complete description of the dynamic evolution. The Hamiltonian for $N$ point vortices located at $z_k=x_k+iy_k$ with a strength $\Gamma_k$ is
$$ H(\xbf) = -{1\over 4 \pi} \,\sum_ {j\not=k} \, \Gamma_j\Gamma_k \log |z_j-z_k| \eqno(76)$$

The coordinates $x_k$ and $y_k$ are canonically conjugated.
\begin{figure}[htb!]
 \centering
\centerline{
\includegraphics[width=8cm,height=7 cm]{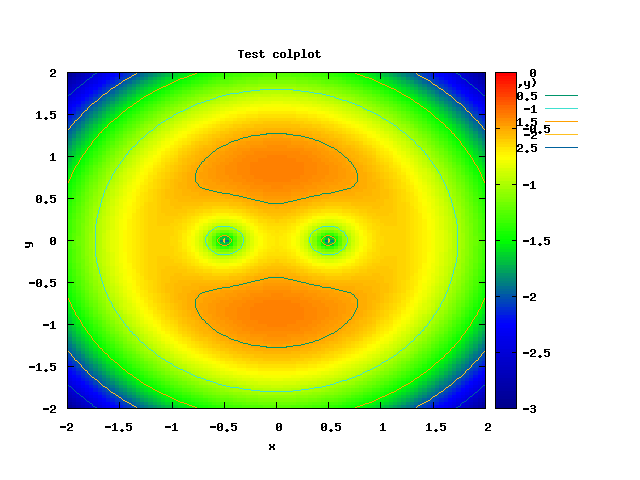}
}
 \caption{\label{Figure 3 } Level lines and isocolor for the three vortices Hamiltonian in the symmetric case $\lambda=1/2$}
\end{figure}

There are two independent first integrals in involution. As a consequence the system of 3 vortices is integrable, while a system of 4 is not .
If the vorticity of one of them is very low, a reduced Hamiltonian can be written which looks similar to the restricted planar three body problem.
Introducing a canonical change of coordinates the Hamiltonian reads
$$ H= -{1\over 2} (x^2+p^2) +(1-\lambda) \log [(x+\lambda)^2+p^2]^{1/2} + \lambda \log [(x+\lambda-1)^2+p^2]^{1/2} \eqno(77) $$

%
% The 4 vertex Hamiltonian with equal vorticities takes the following form
%
% $$ H= -(\log p_1+\log p_2) -4\log \Bigl [ 1+p_1+p_2+\sqrt{p_1}\, \sqrt{1-p_1-p_2} \cos\theta_1 + \Bigr . $$
%
%
% $$ \Bigl . + \sqrt{p_2}\,\sqrt{1-p_1-p_2}\cos\theta_2 +\sqrt{p_1p_2} \cos(\theta_1-\theta_2) \Bigr] \eqno(78) $$
%
%
\
We remark that the splitting cannot be used to construct an explicit symplectic integrator. Only
symplectic implicit schemes are available. However one can use a
4-th order Runge-Kutta integrator choosing $n_s=10^4$ such that the Hamiltonian is conserved
within an error very small with respect to the stochastic perturbation. In Figure 3 we show level curves of the Hamiltonian (77).
The behaviour of the forward and reversibility error follows the power law with exponent $3/2$ in agreement with the theory.

\section{The observational noise}
To complete the present analysis on the effect of noise on a dynamical system we consider now the effect of observations.
We suppose they are made regularly with a time interval $T$. We assume our system to be regular and
to be given by the translation on the torus $\Toro$ and that each observation, having a small duration $\tau\ll T$,
behaves as a white noise of amplitude $\epsilon$.
The correlation and the fidelity for an observable $f_\Toro(x)$ during the first measurement
$T<t<T+\tau$ decay. After this time interval the correlation oscillates, whereas the fidelity remains constant, namely for $t>T+\tau$
is
$$ C_\epsilon(t)=\sum_{k\not=0}\,\, |f_k|^2 e^{2\pi\,i\,\omega \,t}\, e^{-2\pi^2\epsilon^2 \tau}
\qquad \qquad F_\epsilon(t)=\sum_{k\not=0}\,\, |f_k|^2 e^{-2\pi^2\epsilon^2 \tau}
\eqno(78)$$
If we repeat measurements of the same duration $\tau$ and same time separation $T$,
after $n$ of them, namely for $t\geq n(T+\tau)$,
the decrease factor $e^{-2\pi^2\,\epsilon^2\,n\tau}$ becomes significant
if $n\,\epsilon^2\tau >1$. The threshold $t_*$
to observe the exponential decrease is
$$ t_* \sim {T\over \epsilon^2\,\tau} \eqno(79)$$

\subsection{Map formulation}

If the disturbance introduced by the measurement is proportional
to the signal itself, then we must treat it as a multiplicative noise.
In the continuous time case analytical
estimates are difficult to obtain also for translations on the torus.
To overcome the difficulty
we consider the case of a discrete time system.
When the noise is additive the results are similar to the continuous time case.
We consider a map $M$ which gives the evolution on the same time interval $T$
which separates two subsequent observations. Letting $x_{n+1}=M(x_n)$ we have
$$ x_{n}= x_{n-1} +\omega T + \epsilon\, \tau^{1/2} \,\xi_n\,\mod 1 \eqno(80)$$
where $\xi_n$ are random variables uniformly distributed in $[-1,1]$.
The correlation after $n=1$ iteration (observation) is
$$ \eqalign{ C_\epsilon(1) & = \int_0^1 dx_0 \,f_\Toro(x_0) \,
\langle f_{\Toro}(x_0+\omega T+\epsilon \,\tau^{1/2}\xi ) \rangle =\cr
&= \sum_{k,k'\not=(0,0)} \,f_{k}f_{-k'} \, e^{2\pi\,i\,k\,\omega T}\,{1\over 2}\,
\int_{-1}^1\,d\xi\, e^{2\pi\,i\,k \epsilon \,\tau^{1/2}\,\xi}\times \int_0^1 \,dx_0\, e^{2\pi (k'-k)x_0} =\cr
&= \sum_{k\not = } \,|f_{k}|^2\,\,
e^{2\pi\,i\,k\,\omega T}\,S(\epsilon\,\tau^{1/2}\, k)
\qquad \qquad S(x)= {\sin(2\pi x)\over 2\pi x}\cr} \eqno(81) $$
After $n$ iterations (observations) the correlation becomes
$$ C_\epsilon(n)= \sum_{k\not = } \,|f_{k}|^2\,
\,e^{2\pi\,i\,k n\omega T}\,S^n(\epsilon\, \tau^{1/2} k) \eqno(82)$$
The same result holds for the fidelity where the phase factor is absent.
We recall that $|S(x)|\leq e^{- x^2\log(2\pi)}$ for $|x| \leq 1$ and $|S(x)|\leq (2\pi|x|)^{-1}$ for $|x|>1$, see reference \cite{Marie}, and
follow the same procedure outlined there. Assuming $\epsilon\,\tau^{1/2} \ll 1 $ and that the Fourier coefficients
are bounded by $|f_k|\le A/|k|$, we have
$$ |C(n) | \leq 2A^2\,e^{-n\, \tau\, \epsilon^2\,\log(2\pi) } \eqno(83)$$
\subsection{Multiplicative noise}
If the random perturbation introduced by an observation is proportional to the
signal effect and we make a single observation after $N$ iterations the map (82) is replaced by
$$ x_{n}= (x_{n-1} +\omega\,T)(1+\epsilon \,\tau^{1/2} \xi_n) \, \mod 1 \eqno(84)$$
After the first iteration the correlation reads
$$ \eqalign{ C_\epsilon (1) &=
\sum_{k,k'\not=(0,0)} \,f_{k}f_{-k'} \,e^{2\pi\,i\,k\omega T}\, {1\over 2}\,
\int_{-1}^1\,d\xi\, e^{2\pi\,i\,k \,\omega T\, \epsilon \tau^{1/2}\xi} \int_0^1 \,dx_0 \,
e^{2\pi\,i\,(k(1+\epsilon \tau^{1/2}\xi) -k') x_0} = \cr
& = \sum_{k,k'\not=(0,0)} \,f_{k}f_{-k'} (-1)^{k-k'} \,e^{2\pi\,i\,k\omega T}\, {1\over 2}\,
\int_{-1}^1\,d\xi\,S\parton{k-k'+k\epsilon \tau^{1/2}\xi\over 2} \cdot \cr
&\cdot \exp\parton{2\pi \,i\, k\epsilon \tau^{1/2} \xi \parton{\omega T +{1\over 2} } } \cr} \eqno(85) $$
If $f_\Toro(x)$ is a trigonometric polynomial, namely $f_k=0$ for $|k|\geq K$ and $\epsilon \tau^{1/2} K\ll1$
then
$S((k-k'-k\epsilon \tau^{1/2} \xi)/2)\sim \delta_{k,k'}$ so that
$$ C_\epsilon(1) = \sum_{k\not=0} \,|f_k|^2 \,e^{2\pi\,i\,kn\omega T } \, S\parton{k\epsilon \tau^{1/2}\parton{\omega T+{1\over 2}}} \eqno(86)$$
Supposing that $\epsilon\tau^{1/2}\, \omega T \gg 1$ a single observation produces a significant decrease in
the correlation roughly proportional to $(\tau^{1/2}\,\epsilon \,\omega T)^{-1}$.
On the contrary if $ \epsilon\tau^{1/2} \, \omega T \ll 1$ the decrease is negligible and we need to make several observations. After $n$
observations we have

$$ C_\epsilon(n) = \sum_{k\not=0} \,|f_k|^2 (-1)^{k-k'} \,e^{2\pi\,i\,k n\omega T } \, S^n\parton{k\,\tau^{1/2}\epsilon \parton{\omega T+{1\over 2}}} \eqno(87) $$
Supposing that $\tau^{1/2} \epsilon \ll1 $ and $\omega T \gg 1$ we have the following estimates
$$ C_\epsilon(n) \leq 3A^2\cases{ \exp\parton{-\log(2\pi) \,n \tau \epsilon^2 (\omega T)^2} & if $ \tau^{1/2} \epsilon \,\omega T \ll 1$ \cr
& \cr
(2\pi\,\tau^{1/2}\epsilon \, \omega T)^{-n} & if $ \tau^{1/2} \epsilon \,\omega T \gg 1 $ \cr} \eqno(88)$$
To conclude, if the stochastic perturbation, introduced by an observation, is multiplicative rather than
additive, a phase factor $\omega T$ is introduced in the correlation decay. Since the perturbation amplitude and duration are assumed to be small
($\tau^{1/2} \epsilon \ll 1$), if the phase factor is large ($\omega T \gg 1$), the decay is much faster. If 
$ \tau^{1/2}\epsilon\omega T \sim 1 $,
a few observations determine a significant decay of correlations in the multiplicative case.

\section{Conclusions} 
We have examined the asymptotic behaviour of the forward and reversibility errors (FE, RE) introduced by stochastic perturbations
in a dynamical system defined on a compact phase space and we have compared these errors with the decay rates of
correlation and fidelity. %which define the memory loss with respect to the initial condition and to the unperturbed trajectory.
We have first considered the translations on the torus $\Toro$ and the cylinder, showing that 
both errors follow the same power law with an exponent equal to $1/2$ and $3/2$ respectively.
%The forward error FE is defined as the root mean square deviation with respect to the unperturbed orbit, 
%the reversibility RE as the mean square deviation with respect to the reversed orbit.
For expanding flows on $\Toro$ the FE grows exponentially, but the equivalence of FE and RE
occurs only if we have a linear flow on the torus $T^2$ which preserves the area. 
This is a special case of linear Hamiltonian flows for which the asymptotic equivalence of FE and RE is proved. 
We have analysed the relation between the memory loss and the error growth.
%To this end we have considered the correlations and fidelity.
For linear Hamiltonian systems we have shown that the correlation and fidelity decay as $e^{-2\pi^2\sigma^2(t)}$
where $\sigma(t)$ is the FE and that a similar relation holds between the fidelity and the RE.
The RE method can be used to investigate the transition regions
from regular to chaotic motions and the effect of round off errors in
numerical simulations. The proposed framework
%is very general and could also be extended to field equations having a Hamiltonian structure,
%though not canonical, as it is the case for the wave equations of an inviscid fluid.
is general and an extension to classical fields, by using the theory of linear stochastic partial differential
equations to define the FE and RE, could be considered. 
%Their asymptotic equivalence can be expected only for the field equations having a Hamiltonian structure, as in the case of inviscid fluids.
Finally we have examined the effect of the observational noise showing that for an integrable system only a sequence of observations
can lead to the decay of the correlation and that the multiplicative noise can increase the decay rate if the phase advance
between two observations is large.

\section*{Acknowledgements}
GT and SV would like to thank the Newton Institute of Cambridge for the kind hospitality during the semester Mathematics for Planet Earth, during which this paper was initiated. SV was supported by the Leverhulme Trust thorough the Network Grant IN-2014-021 and by the MATH AM-Sud Project ``Physeco''. SV also thanks the Department of Physics of the University of Bologna for the kind hospitality during the completion of this work.

\appendix
\section{Correlation on the torus and Perron Frobenius operator}
\label{app1}
Given a map $M(x)$ continuous on the torus there is a unique definition of correlation. Indeed the evolution of a given
initial distribution $f_\Toro(x)$ is governed by the Perron Frobenius equation
$$ f_\Toro(x,n)= (\Pop^n f_\Toro)(x)=\int_0^1 \, \delta (x-M^n(x_0)) f_{\Toro} (x_0) dx_0 = \sum_{j=1}^{q^n} {f_{\Toro} (x_j)\over (M^n)'(x_j)}\eqno(A1) $$
where $\Pop$ is the Perron-Frobenius operator. Supposing $M(x_0)$ has $q$ preimages,
with $x_j=(M^{-n})_j(x)$ for $1\leq j\leq q^n$ we have indicated all the preimages
of $M^n(x_0)$.
For any $A\subset \Toro^1$ the
invariance of the probability measure $\mu$ is expressed by $\mu(M^{-1}(A))=\mu(A)$.
For example the Bernoulli map $M(x)=q x\,\mod 1$ where $q\in \Naturali$
has $q$ inverses and $M^{-1} \equiv \cup_k M^{-1}_k(A)$ where $\mu(M^{-1}_k(A))=q^{-1}\mu(A)$.
Since the density of the absolutely continuous invariant measure is 1, we define the correlation
according to
$$ \eqalign{ C(n) &= \int_0^1 \,f_\Toro(x_0) f_\Toro(M^n(x_0)) \,dx_0 -f_0^2 = \cr
& =\int_0^1 \,dx_0\, f_\Toro(x_0) \int_0^1 \delta(x- M^n(x_0)) \,f_\Toro(x)\,dx-f_0^2 = \cr
& =\int_0^1 \, f_\Toro(x) f_\Toro(x,n) \,dx -f_0^2 \cr } \eqno(A2)$$
where the result has been obtained by interchanging the integration order.

As a specific example we consider the
linear map $M(x)=qx \,\mod 1$ continuous on $\Toro$ for integer $q$.
In this case the preimages $x_j$ of $M^n$ are
$$ x_j= (M^{-n})_j(x)= {x\over q^n}+ {j_n\over q}+ {j_{n-1}\over q^2}+\ldots+ {j_2
\over q^{n-1}}+ {j_1\over q^n}= {x+j\over q^n}
\eqno(A3)$$
where $j_i=0,1,\ldots,q-1$ and $j=j_1+ q j_2+\ldots + q ^{n-2} \,j_{n-1}+ q^{n-1} j_n$ has a range from $0$ to $q^n-1$.
As a consequence, 
%$f_\Toro(x,n)= \sum_{j=0}^{q^n-1}\, q^{-n}\,f_\Toro(q^{-n}(x+j) $
from the Perron-Frobenius equation we have
$$ f_\Toro(x,n)= \sum_{j=0}^{q^n-1}\, q^{-n}\,f_\Toro(q^{-n}(x+j) \eqno(A4)$$
and the following equality can be proved
$$ \eqalign{ & \int_0^1 \,f_\Toro(x,n)\,f_\Toro(x) \,dx = \sum_0^{q^n-1}\, \int_0^1 \, {1\over q^n}\,f_\Toro\parton{x+j\over q^n}\,f_\Toro(x) = \cr
& = \sum_{j=0}^{q^n-1} \int_{j/q^n}^{(j+1)/q^n} \, dx_0\, f_\Toro(x_0) \,f_\Toro\parton{q^n \, x_0 -j}\,
=\int_0^1 \,dx_0 f_\Toro(x_0)\,f_\Toro(q^n x_0) \cr } \eqno(A5) $$
where we have made the coordinates change $x+j=q^n x_0$. Letting $f_k$ be the Fourier coefficients of $f_\Toro(x)$ the order $k$
Fourier coefficients of $f_\Toro(x,n)$ are $f_k(n)=f_{q^n k}$.
%and of $f_\Toro(q^nx)$ are the same $f_k(n)=f_{q^n k}$ (recall that $f_k(n)$ is defined for
%$k$ integer but $f_k$ is an analytic function of $k$).
The equality (A4) is a consequence of continuity of the map $M$ on $\Toro$. For a linear flow $S_t(x_0)=e^{\beta t}\,x_0$
we have seen that a similar equality, given by equation (22),
holds only for the sequence of values of $t_m$ such that $e^{\beta t_m} =m$ is an integer, since the map
$S_{t_m}(x_0)= m\,x_0 \,\hbox{mod} \, 1$ is continuous on $\Toro$.

%\section*{References}
%\bibliographystyle{iopart}
%\bibliography{bibliografia}

\end{document}